\documentclass[apj]{emulateapj}

\usepackage{amsmath, amsfonts, amssymb, fancyhdr}
\usepackage{natbib}

\newcommand{\nhtwo}{n({\rm H_2})}
\newcommand{\cmc}{\;{\rm cm}^{-3}}
\newcommand{\cms}{\;{\rm cm}^{-2}}
\newcommand{\gcms}{\;{\rm g\;cm}^{-2}}
\newcommand{\kms}{\;{\rm km\;s}^{-1}}

\newcommand{\mcol}[1]{\multicolumn{1}{c}{#1}}

\newcommand{\boh}{B_{\rm OH}}
\newcommand{\bism}{B_{\rm ISM}}
\newcommand{\bmin}{B_{\rm min}}
\newcommand{\beq}{B_{\rm hyd}}
\newcommand{\bsnr}{B_{\rm SNR}}

\begin{document}

\title{The Role of Magnetic Fields in Starburst Galaxies as Revealed by OH Megamasers}
\author{James McBride, Eliot Quataert, Carl Heiles, and Amber Bauermeister}
\affil{Department of Astronomy, University of California, Berkeley, CA 94720-3411;
    \\jmcbride@astro.berkeley.edu, eliot@astro.berkeley.edu, heiles@astro.berkeley.edu
}

\begin{abstract}
    We present estimates of magnetic field strengths in the interstellar media of starburst galaxies derived from measurements of Zeeman splitting associated with OH megamasers. The results for eight galaxies with Zeeman detections suggest that the magnetic energy density in the interstellar medium of starburst galaxies is comparable to their hydrostatic gas pressure, as in the Milky Way. We discuss the significant uncertainties in this conclusion, and possible measurements that could reduce these uncertainties. We also compare the Zeeman splitting derived magnetic field estimates to magnetic field strengths estimated using synchrotron fluxes and assuming that the magnetic field and cosmic rays have comparable energy densities, known as the ``minimum energy'' argument. We find that the minimum energy argument systematically underestimates magnetic fields in starburst galaxies, and that the conditions that would be required to produce agreement between the minimum energy estimate and the Zeeman derived estimate of interstellar medium magnetic fields are implausible. The conclusion that magnetic fields in starburst galaxies exceed the minimum energy magnetic fields is consistent with starburst galaxies adhering to the linearity of the FIR-radio correlation.

\end{abstract}

\section{Introduction} \label{intro}
The magnetic field in the Milky Way plays a dynamically important role in the interstellar medium; the magnetic field, cosmic rays, and turbulent motion each provides a comparable contribution to pressure support for interstellar matter in the Galactic plane \citep{Boulares1990}. Magnetic fields are also important for redistributing angular momentum, allowing gas clouds to collapse to form molecular clouds and star clusters \citep[][]{Kim2003}. Magnetic field strengths in other star forming galaxies are often estimated assuming the same rough equality between cosmic ray energy density and magnetic field energy density. In this manner, measurements of the synchrotron flux and the volume of the synchrotron emitting region can be used to estimate magnetic field strengths, an argument first laid out by \citet{Burbidge1956}. This argument applied to nearby star forming galaxies suggests that they contain magnetic fields with strengths comparable to the Milky Way \citep{Fletcher2010}, and that magnetic fields generally play a dynamically important role in star forming galaxies \citep{Beck2013}. 

\citet[][hereafter T06]{Thompson2006} distinguished the ``minimum energy'' field, $\bmin$, in which the cosmic ray and magnetic field energy densities are comparable, from the ``equipartition'' magnetic field, in which magnetic fields have energy densities comparable to the hydrostatic pressure, and which we will call $\beq$ (T06 used $B_{\rm eq}$ instead). They found that while $\beq \sim \bmin$ for typical star forming galaxies, $\beq \gg \bmin$ in starburst galaxies. If the true volume averaged interstellar medium (ISM) magnetic field strength in starburst galaxies, which we call $\bism$, is comparable to $\bmin$, then magnetic fields in starbursts are dynamically unimportant. There is no physical reason, however, to expect the assumptions made in the minimum energy argument to hold in all cases. This is particularly true when $\beq \gg \bmin$, as in that case field strengths in excess of $\bmin$ are dynamically reasonable. Indeed, T06 argued that $\bism$ likely exceeds $\bmin$.

Their argument was based in part on the observed correlation between the far infrared (FIR) and radio luminosities of star forming galaxies, which extends over 4 orders of magnitude in luminosity, from dwarf galaxies to starburst galaxies \citep[][]{Yun2001}. The FIR-radio correlation has been explained via the ``calorimeter'' theory, in which both the radio and FIR luminosities effectively measure instantaneous star formation rates \citep[V\"{o}lk 1989; see][for a detailed discussion, and references to alternative explanations]{Lacki2010a}. The FIR emission comes from dust heated by massive young stars, while the radio emission results from relativistic electrons produced in supernovae. 

T06 highlighted an apparent inconsistency between the FIR-radio correlation and the minimum energy magnetic fields in starburst galaxies. The synchrotron cooling time, which scales as $\tau_{\rm syn} \propto B^{-3/2}$, is longer than the inverse Compton cooling time if $\bism = \bmin$. The minimum energy magnetic field strengths also suggest synchrotron cooling times for cosmic ray electrons that are longer than their escape time (set, for example, by advection in a galactic wind). If the relativistic electrons produced in supernovae do not primarily lose their energy via synchrotron emission, then there would be no reason to expect the radio luminosity to be proportional to the FIR luminosity in starburst galaxies. No deficit in radio luminosity is observed, however, so T06 argue that the minimum energy argument underestimates the true magnetic field strength in starburst galaxies.

Our aim is to test the prediction of T06 that $\bism \gg \bmin$ by estimating magnetic field strengths in starburst galaxies independent of the minimum energy argument. We developed a method for estimating $\bism$ that relies upon Zeeman splitting associated with OH megamasers, as described in Section \ref{sec:data}. In Section \ref{sec:results}, we present the estimates of $\bism$ made using this method, compare the estimated fields to the minimum energy magnetic fields, and argue that the minimum energy argument does underestimate magnetic field strengths in starburst galaxies. In Section \ref{sec:discussion}, we discuss the implications of this result for the FIR-radio correlation and the dynamic importance of magnetic fields in the ISM of starburst galaxies. Finally, in Section \ref{sec:conclusions} we provide a summary of our main conclusions.

\section{Data} \label{sec:data}

\begin{table*}
    \centering
    \caption{List of magnetic field measures discussed} \label{tab:mag_def}
    \begin{tabular}{l l}
        Symbol & Description \\ \hline
        $\bism$ & Volume averaged magnitude of the magnetic field in the starburst ISM \\
        $\boh$ & Line-of-sight magnetic field measured in OH masing clouds \\
        $\bmin$ & Magnetic field estimated from synchrotron flux, assuming comparable magnetic field and cosmic ray energy density \\
        $\beq$ & Magnetic field required to balance the disk against its own self-gravity \\
        $\bsnr$ & Magnetic field strength in ISM estimated from synchrotron in supernova remnants (see \citealt{Thompson2009})
    \end{tabular}
\end{table*}
Testing the dynamic importance of magnetic fields in starbursts requires a constraint on the gas surface densities in the starburst nucleus and a method of estimating magnetic field strengths in the ISM of starburst galaxies that does not rely on the minimum energy argument. The basic procedure we used to do this is as follows:
\begin{enumerate}
    \item Select a sample of starburst galaxies with measurements of Zeeman splitting associated with OH masers and in which the CO(1-0) line has been detected
    \item Assume $\nhtwo \sim 10^4 \cmc$ is typical in OH masing clouds \citep{Lockett2008}
    \item Estimate the molecular gas number density and surface density using the intensity and width of the CO(1-0) line and the dust temperature derived from IRAS, following \citet[][hereafter S97]{Solomon1997}
    \item Assume $\nhtwo \sim 10^3 \cmc$, which is roughly the median of estimates from step 3, is typical in the ISM of starburst galaxies 
    \item Assume that $B \propto n^{1/2}$, and then use the Zeeman derived magnetic field strength and the densities in the masing regions and the ISM to arrive at an estimate for the typical magnetic field in the ISM
\end{enumerate}
These steps are described in greater detail in the following two subsections. The different magnetic fields discussed are summarized in Table \ref{tab:mag_def}.

\subsection{Magnetic fields in OH masing clouds} \label{sec:mag_data}
We use measurements of Zeeman splitting associated with OH maser lines as a starting point for estimating magnetic field strengths in the ISM of starburst galaxies. Some (ultra)luminous infrared galaxies ([U]LIRGs) host powerful OH masers, called OH megamasers (OHMs), with roughly $1/3$ of the warmest ULIRGs hosting OHMs \citep{Darling2002a}. Recent work suggests that the (U)LIRGs that host OHMs are predominantly starburst, rather than AGN, dominated \citep{Willett2011a, Willett2011}.  The sample of Zeeman splitting of OH masing lines is taken from \citet{Robishaw2008} and \citet{McBride2013}. They made observations primarily using Arecibo Observatory, though one source used here was observed with Green Bank Telescope, both of which are single dish telescopes. As a result, the OH masers are not spatially resolved in the Zeeman observations. 

Zeeman splitting is most easily observed in bright, narrow maser lines, and typical Zeeman splitting detections were associated with lines that have velocity widths $\lesssim 20\kms$ and flux densities $\gtrsim 3\;$mJy. The magnetic fields derived from measurements of Zeeman splitting, which we call $\boh$, are likely to be at least somewhat larger than $\bism$, as the maser lines for which Zeeman splitting can be measured likely arise in regions that are smaller and denser than the starburst region of (U)LIRGs. In Sections \ref{sec:cloud_size} and \ref{sec:cloud_density}, we discuss the conditions in the regions of the ISM where the maser lines with Zeeman splitting are produced, for the purpose of understanding how $\boh$ relates to $\bism$.

\subsubsection{Masing cloud sizes} \label{sec:cloud_size}
The synchrotron emitting region in (U)LIRGs has a size scale of $\sim$100--1000~pc \citep{Condon1991}. There are two reasons to believe that the regions that give rise to the bright, narrow lines for which Zeeman splitting can be detected have typical sizes of order $\sim$pc, meaning that $\boh$ is not directly probing the field in the starburst region as a whole. 

The most direct evidence for compact masing clouds is high resolution observations that have been done for a sample of the brightest OHMs. Very Long Baseline Interferometry (VLBI) observations of three nearby OHMs reveal emission on scales of $\sim$1--10~pc (Arp~220: \citealt{Lonsdale1998}; III~Zw~35: \citealt{Diamond1999, Pihlstrom2001}; IRAS~F17207--0014: \citealt{Diamond1999, Momjian2006}). Many of the compact regions that appear in VLBI maps correspond to the bright, narrow features in which Zeeman splitting can be detected in single dish observations. Additional OHMs have been observed with physical resolutions of $\sim$50--100~pc. In Markarian~273 \citep{Yates2000} and IRAS~F12032+1707 \citep{Pihlstrom2005}, most emission occurs on scales $\lesssim$100~pc. Compact emission is absent in Markarian~231 \citep{Klockner2003,Lonsdale2003}, though \citet{Richards2005} interpreted their observations of OHM emission in Markarian~231 as consistent with a torus inclined to the plane of the sky at $\sim$45$^\circ$ that had individual masing clouds with sizes $\sim$1~pc. The absence of bright, compact features is then a consequence of limited cloud overlap. Finally, VLBI observations of IRAS~F14070+0725 \citep{Pihlstrom2005} resolved out most OHM emission, from which they conclude that much of the emission is diffuse. Altogether, however, the VLBI observations show that in most OHMs, bright, narrow maser lines are produced in regions with sizes $\sim$pc. 

Variability in OHMs provides additional evidence that narrow lines correspond to compact masing regions. \citet{Darling2002b} reported the first evidence of variability in an OHM. They observed IRAS~F21272+2514 over a time baseline of more than two years. They found rms variation of $\sim$10\% in the two brightest components, with variation on timescales at least as short as 39 days, which was the shortest interval between observing epochs. If the variability was intrinsic to the maser, this would suggest a source size $\lesssim$0.03~pc. \citet{Darling2002b} instead interpret the variability as being produced by interstellar scintillation primarily from the electron screen in the Milky Way. In this interpretation, roughly half of this OHM's emission is produced in regions with sizes $\lesssim$2~pc, with the remainder coming from larger regions. 

There has not been a similarly detailed study of variability in OHMs since, but there is nonetheless evidence that this type of variation is common in OHMs. \citet{Lonsdale2008} saw variability in Arp~220, and \citet{Robishaw2008} noted a factor of $\sim$2 change in the flux of the brightest component in IRAS~F12032+1707, a component for which they also measured an 18~mG magnetic field. \citet{McBride2013} examined spectra for six sources observed in 2006 and presented in \citet{Robishaw2008} that were observed again in 2008--2009, and found variation at the 5-10\% level in four of the sources. These observations of variability qualitatively suggest that the results found by \citet{Darling2002b} are more generally representative of OHMs.

VLBI observations and variability in OHMs indicate that magnetic field strengths derived from Zeeman splitting of OH maser lines are associated with regions that have typical sizes $\lesssim$1--10~pc. Thus they are smaller, and likely denser, than the starburst as a whole, which means that $\boh$ is likely to exceed $\bism$. We argue that $\boh$ is not likely to be significantly larger than $\bism$, however, given the densities of masing regions relative to the ISM of starburst galaxies, and that $\boh$ can be reasonably scaled to provide an estimate of $\bism$.

\subsubsection{Masing cloud gas volume density}  \label{sec:cloud_density}
There are not direct measurements of the density of compact masing clouds in which the bright, narrow maser lines are produced.  
Instead, we rely upon OHM modeling results that have been generally successful at explaining OHM emission. Early attempts concluded that OHM emission could be produced via radiative pumping and gas with densities $\nhtwo \sim 10^3$--$10^4 \cmc$ \citep{Burdyuzha1990a, Randell1995}. The subsequent discovery of compact features in OHMs seemed to suggest unrealistically high radiative pump efficiencies, and led to consideration of collisional pumping and higher densities \citep{Lonsdale2002}. However, \citet{Parra2005a} geometrically modelled III~Zw~35, a source with compact features, and explained its emission in terms of a clumpy ring, in which masing clouds had typical densities $\nhtwo \sim 3 \times 10^3 \cmc$. \citet{Momjian2006} observed IRAS~F17207--0014 and found masing characteristics with remarkable similarity to those of III~Zw~35, and they suggested that a model similar to that of \citet{Parra2005a} could explain the emission they observed. 

\citet{Lockett2008} performed the first detailed pumping calculations of OHMs after the discovery of compact emission in OHMs. They found that radiatively pumped gas with a density $\nhtwo \sim 10^4 \cmc$ could produce the general features of OHMs, including the observed mixture of diffuse and compact emission, and the relative strengths of the 1667~MHz and 1665~MHz emission. Their model also accounted for the satellite lines at 1612~MHz and 1720~MHz being significantly weaker than the 1667~MHz line in the small number of OHMs in which the satellite lines had been observed. The general weakness of satellite line emission in OHMs has since been confirmed in a survey of all known OHMs accessible to Arecibo \citep{McBride2013a}. 

\citet{Lockett2008} also found that the efficiency of masing began to decrease at densities above $\nhtwo \sim 10^4 \cmc$. Though gas at densities above and below $\nhtwo \sim 10^4 \cmc$ will contribute to amplifying maser lines, the exponential nature of maser amplification means that gas with the highest masing efficiency will dominate amplification. Likewise, Zeeman splitting associated with masing lines will receive contributions from gas at different densities, but primarily will reflect the magnetic field strength in the gas where the majority of amplification occurred. {\em Hereafter, we assume that magnetic fields measured via Zeeman splitting in OHM lines reflect gas with densities of $\nhtwo \sim 10^4 \cmc$.}

\begin{table*}
    \caption{Magnetic field data} \label{tab:data}
    \centering
    \begin{tabular}{l l r r r r r r r r r r r r r} \hline \hline
        IRAS Name & Other Name & z & \mcol{$S_{\rm CO}$} & \mcol{$\Delta V_{\rm CO}$} & \mcol{$S_{1.4}$} & \mcol{$S_{60}$} & \mcol{$S_{100}$} & \mcol{$\Sigma_g$} & \mcol{$\boh$} & \mcol{$\bism$} & \mcol{$\bmin$} & \mcol{$\beq$} & CO ref.\\
                  & &  & \mcol{Jy$\kms$} & \mcol{$\kms$} & \mcol{mJy} & \mcol{Jy} & \mcol{Jy} & \mcol{g$\cms$} & \mcol{mG} & \mcol{mG} & \mcol{mG} & \mcol{mG} & \\ \hline
        F01417+1651   & III Zw 35  & 0.027719   & 49.0   & 160   & 40.6   & 12.6   & 13.31   & 1.4   & 2.9   & 1.0   & 0.22   & 3.2 & 1  \\ 
        F04332+0209   &  & 0.012014   & 5.8   & 54   & 4.3   & 3.4   & 3.8   & 0.7   & 46.0   & 15.3   & 0.23   & 1.5  & * \\ 
        F08071+0509   &  & 0.053463   & 71.0   & 230   & 36.3   & 4.5   & 6.9   & 1.3   & $<$20.0   & $<$6.7   & 0.16   & 3.0 & 2  \\ 
        F10039--3338   &  & 0.0341   & 65.0   & 150   & 10.3   & 8.9   & 8.0   & 0.9   & 1.8   & 0.6   & 0.13   & 2.1  & 3,4 \\ 
        F12032+1707   &  & 0.21779   & 3.4   & 240   & 28.7   & 1.4   & 1.5   & 1.9   & 15.0   & 5.0   & 0.46   & 4.3  & 5 \\ 
        F12112+0305   &  & 0.073   & 48.0   & 200   & 23.8   & 8.5   & 10.0   & 0.9   & $<$3.0   & $<$1.0   & 0.16   & 2.1 & 2   \\ 
        F12243--0036   &  & 0.007048   & 165.0   & 130   & 41.4   & 40.7   & 32.8   & 1.8   & $<$20.0   & $<$6.7   & 0.19   & 4.0 & 2   \\ 
        F13126+2453   &  & 0.013049   & 55.0   & 230   & 31.2   & 17.9   & 18.13   & 8.2   & $<$30.0   & $<$10.0   & 0.30   & 18.8 & * \\ 
        F14070+0525   &  & 0.265243   & 3.6   & 270   & 4.0   & 1.5   & 1.8   & 2.0   & $<$16.0   & $<$5.3   & 0.25   & 4.5 & 6 \\ 
        F15107+0724   &  & 0.012705   & 118.0   & 230   & 53.8   & 20.8   & 29.4   & 4.5   & $<$12.0   & $<$4.0   & 0.22   & 10.4 & 2  \\ 
        F15327+2340   & Arp 220  & 0.018116   & 491.0   & 360   & 326.8   & 103.8   & 112.4   & 5.3   & 2.8   & 0.9   & 0.24   & 12.1 & 6   \\ 
        F18368+3549   &  & 0.11617   & 15.0   & 330   & 21.0   & 2.2   & 3.8   & 3.2   & 22.0   & 7.3   & 0.26   & 7.3  & 6 \\ 
        F18588+3517   &  & 0.10665   & 21.0   & 750   & 5.9   & 1.5   & 1.8   & 25.5   & 16.0   & 5.3   & 0.25   & 58.5 & *  \\ 
        F20550+1655   &  & 0.036125   & 80.0   & 180   & 43.9   & 13.3   & 10.6   & 1.3   & 18.0   & 6.0   & 0.20   & 2.9 & 2  \\ 
    \end{tabular}
    \tablecomments{
        Redshifts are taken from SIMBAD. FIR fluxes come from IRAS. Fluxes at 1.4 GHz are from the NRAO VLA Sky Survey \citep{Condon1998}. $\bism$ is simply $\boh / 3$. Calculations assume the {\em WMAP}+BAO+$H_0$ cosmology from \citet{Komatsu2011}. \\
        References for CO(1-0) data: (1) \citealt{Sanders1991}, (2) \citealt{Baan2008}, (3) \citealt{Leech2010}, (4) \citealt{Papadopoulos2012}, (5) \citealt{Combes2011}, (6) \citealt{Solomon1997}, (*) Previously unpublished data acquired with CARMA
    }
\end{table*}

\subsection{Starburst ISM} \label{sec:ism_data}
In a medium with highly supersonic motions---as is the case in (U)LIRGs, given the large linewidths of molecular tracers---gas will occupy a wide range of densities. The density of gas occupying most of the volume can be smaller than the mean density ($\langle n \rangle = {\rm mass} / {\rm volume}$), with volume averaged densities of $\sim 0.01-0.1 \langle n \rangle$ depending on Mach number. Likewise, the density of gas constituting most of the mass is larger than the mean density by a factor 10--100 \citep{Lemaster2008}. For comparison of $\bism$, $\bmin$, and $\boh$, we are interested in estimating the density of gas occupying most of the volume in typical OHM hosts.

Detailed, high-resolution observations of OHM hosts that can provide direct information about their ISM conditions are limited. Among the sample of galaxies with detections of Zeeman splitting, only Arp~220 has an interferometric CO map with a clearly resolved disk \citep{Downes1998}. An additional galaxy with an interferometric CO map, IRAS~F15107+0724, is marginally resolved \citep{Planesas1991}, and has only an upper limit $\boh < 12$~mG on magnetic field strength associated with the masing region. Moreover, observations of many molecular lines in galaxies like Arp~220 indicate that CO(1-0) traces a low-density phase of the molecular ISM \citep{Greve2009}. For this reason, we begin with a discussion of molecular line observations of Arp~220, which is the most well studied source in the sample and is close to the median of the sample as measured by its $L_{\rm CO}$, $L_{\rm FIR}$, and $L_{\rm 1.4\; GHz}$. We then discuss the results of determining volume and gas surface densities individually for each galaxy using the procedure described in S97, and argue that the results for Arp~220 provide a representative model for the rest of our sample.

\subsubsection{Arp 220} \label{sec:arp_data}
Interferometric CO observations of nearby (U)LIRGs like Arp~220 have revealed molecular gas with properties significantly different in the Milky Way. In particular, CO in (U)LIRGs fills a significant fraction of the volume in which it is observed, rather than existing primarily in gravitationally bound clouds \citep{Downes1993}, with CO emission arising in a region with $R\sim$~100--1000~pc \citep{Downes1998}. S97 compared the ISM of galaxies like Arp~220 to scaled up versions of normal galactic disks, with CO emanating from a low density, volume filling gas of density $\nhtwo \sim 10^3 \cmc$, and higher density tracers like HCN tracing self gravitating clouds at higher density within the region. Using detailed dynamical models to interpret interferometric CO(1-0) and CO(2-1) line observations, \citet{Downes1998} concluded that $\langle \nhtwo \rangle \sim 10^4 \cmc$ for the $<$100~pc region around each of the two nuclei in Arp~220, while the $\sim$500~pc disk has $\langle \nhtwo \rangle \sim 10^3 \cmc$. As the synchrotron emitting size of Arp~220 is $\sim$100~pc \citep{Condon1991}, the results for the regions immediately surrounding the nucleus are likely more relevant.

Observations of other molecules with higher critical densities find that a significant fraction of gas in Arp~220 is in denser regions. \citet{Cernicharo2006} detected 183~GHz H$_2$O masers in Arp~220, while no 22~GHz H$_2$O masing has been found. Their models reproduced these line strengths for densities $\nhtwo \sim 10^5$--$10^6 \cmc$ in the central $\sim$kpc of Arp~220. \citet{Greve2009} observe many molecular transitions in Arp~220, and argue that molecular gas with density $\nhtwo \sim 10^5$--$10^6 \cmc$ accounts for the majority of the molecular mass, and is concentrated around the nuclei of Arp~220. Their large velocity gradient modeling of observations of different molecular lines suggest two phases of molecular gas: a diffuse phase with $\nhtwo \sim 10^2$--$10^3 \cmc$, and gas with densities of $\nhtwo \sim 10^5$--$10^6 \cmc$ that dominates the mass. These conditions, while very unlike those in normal star forming galaxies, are likely representative of the OHM sample as a whole. For instance, \citet{Papadopoulos2012a} find that similar two-phase models are necessary to explain the CO spectral line energy distributions from a larger sample of (U)LIRGs, and note that such two-phase models are consistent with the highly turbulent ISM in (U)LIRGs. Moreover, \citet{Papadopoulos2012a} argue that using low-$J$ CO lines alone leads to an underestimate of molecular gas mass in LIRGs, a factor we consider in estimating volume averaged gas densities for galaxies in our sample.

In combination, these models and observations suggest that $\nhtwo \sim 10^3 \cmc$ is a reasonable representative density for the gas that occupies most of the volume in the synchrotron emitting region in Arp~220. This value is equal to $0.1 \langle \nhtwo \rangle$, as determined from CO(1-0) observations of the gas on size scales comparable to those on which synchrotron emission is produced. The value $\nhtwo \sim 10^3 \cmc$ is also consistent with conditions estimated from volume filling gas, such as NH$_3$, and is a factor of $\sim$1000 less dense than conditions probed by molecular line tracers such as HCN. We believe that this estimate for the volume averaged gas density is uncertain at the factor of $\sim 10$ level; we take this into account below when providing an estimate of the systematic error in our ISM magnetic field inferences.

\subsubsection{Blackbody model} \label{sec:solomon_data}
As similarly detailed observations do not exist for the other galaxies in our sample, we instead follow the argument laid out by \citet{Downes1993} and S97 to estimate volume and gas surface densities, and compare derived values to those of Arp~220. For (U)LIRGs that do have interferometric CO maps, such as Arp~220, observed H$_2$ column densities are of order $10^{24} \cms$. This column density is large enough to produce an optical depth $\tau > 1$ for 100\;$\mu$m emission. Thus 60 and 100 $\mu$m fluxes measured by IRAS in (U)LIRGs are produced by blackbody dust emission. The flux of CO(1-0) emission, which is optically thick, is well correlated with the 100\;$\mu$m flux, which S97 note also supports the argument that 100\;$\mu$m fluxes are optically thick. Blackbody dust emission is characteristic of OHMs more generally, as IRAS fluxes for OHMs are close to blackbody emission between 25 and 100 $\mu$m \citep{Baan1987,Chen2007}, and mid-IR {\em Spitzer} spectra of OHMs have features consistent with blackbody emission beyond 100~$\mu$m \citep{Willett2011}.

S97 use their blackbody model to provide an estimate for the minimum CO radius in (U)LIRGs in terms of 60 and 100 $\mu$m fluxes, and the luminosity and width of the CO(1-0) line, with
\begin{align}
    R_{\rm CO}({\rm min}) & = 160 \left(\frac{L^{'}_{\rm CO}}{\rm 10^9\; K\; \kms \;pc^2}\right)^{1/2} \times \notag \\
                          & \left(\frac{T_{\rm bb}}{\rm 40\;K}\right)^{-1/2} \left(\frac{\Delta V_{\rm CO}}{300\; \kms}\right)^{-1/2} {\rm pc}. \label{eq:rco}
\end{align}
The blackbody temperature $T_{\rm bb}$ is determined from Planck's law and the IRAS fluxes, with the approximation
\begin{equation}
    T_{\rm bb} \simeq -(1 + z)\left[\frac{82}{\ln(0.3 \frac{S_{60}}{S_{100}})} - 0.5\right] {\rm K},
\end{equation}
where $z$ is the redshift, and $S_{60}$ and $S_{100}$ are the 60 and 100\;$\mu$m fluxes, respectively. S97 then compute a dynamical mass within the CO disk radius, and use a partial correction for inclination effects by taking the maximum of the observed line width and $300 \kms$, finding 
\begin{align}
    M_{\rm dyn}(<R_{\rm CO}) & = 2.1 \times 10^9 \left(\frac{R_{\rm CO}}{{\rm 100\;pc}}\right) \times \notag \\
                             & \left[\frac{{\rm max}(300, \Delta V_{\rm CO})}{300 \kms}\right]^2 M_\odot. \label{eq:dyn_mass}
\end{align}
The values of $R_{\rm CO}$ derived in this fashion are in the range 50--500~pc, which is comparable to the size of the radio emission associated with compact starbursts in (U)LIRGs \citep{Condon1991}, and the size scale where most dust obscuration likely occurs \citep{Murphy2013}. For this reason, we expect $R_{\rm CO}$ to be an appropriate value to take, despite being a lower limit. The same is true for the dynamical mass, which is simply proportional to $R_{\rm CO}$. We then finally calculate a gas surface density, taking simply the dynamical mass divided by the CO disk area, $\pi R_{\rm CO}^2$. Given the above, gas surface densities derived in this fashion are upper limits, as the gas mass will necessarily be less than or equal to the dynamical mass. S97 note though that the gas masses represent a significant fraction of the dynamical mass, so the densities derived in this fashion are likely reasonable estimates.  

We test this for four galaxies in the sample that do have interferometric CO(1-0) observations, and find that the estimates are, in fact, reasonable. For the two aforementioned galaxies in the literature, IRAS~F15107+0724 and Arp~220, the S97 method yields estimates of $\langle \nhtwo \rangle \sim 10^3 \cmc$, which are within a factor of $\sim$2--3 of $\langle \nhtwo \rangle \cmc$ estimates using $\sim$kpc resolution observations. We note, however, that \citet{Downes1998} found $\langle \nhtwo \rangle$ an order of magnitude higher using CO(1-0) emission on the $\sim$100~pc scale on which synchrotron emission is produced in Arp~220. We also used the Combined Array for Millimeter-wave Astronomy (CARMA) to place constraints on the gas surface densities in two more galaxies, IRAS~F04332+0209 and IRAS~F13126+2453. The lower limits on gas surface density from the CO(1-0) maps are comparable to the estimate using the S97 method.

Using this method, gas surface densities $\Sigma_g \sim$ 1--10$\gcms$ and number densities of $\langle \nhtwo \rangle \sim 10^3$--$10^4 \cmc$ appear reasonable for all of the galaxies considered here, and Arp~220 falls near the middle of the range of values. As discussed in the previous subsection, in a turbulent medium, the density of gas that occupies the majority of the volume within a given region will be lower than the average density of that region. However, the observations and modeling of (U)LIRGs by \citet{Papadopoulos2012a} suggest that using the CO(1-0) line leads to an underestimate of the total molecular mass. Moreover, on the scale on which synchrotron emission is produced in Arp~220, the average density exceeds the S97 estimate by an order of magnitude. As the size of the synchrotron emitting region in Arp~220 is typical of similar starburst galaxies \citep{Condon1991}, and the S97 method gives comparable average gas densities on larger scales within Arp~220 as it does for our sample, we consider the results of the previous subsection to be broadly applicable to our sample as a whole. {\em For the remainder of the paper, we take gas surface densities found using the procedure described in S97, and assume the ISM density that fills most of the volume in the region of synchrotron emission is simply $\nhtwo \sim 10^3 \cmc$.}

\subsection{Synthesis} 

With the assumptions discussed in the previous two subsections, we use the 
following procedure. 
We compile a sample of starburst galaxies with Zeeman splitting measurements of OHMs, taken from \citet{Robishaw2008} and \citet{McBride2013}. With one possible exception, the detections and upper limits used here are broadly representative of the observed OHM sample as a whole. No significant differences in galaxy scale properties were found between the galaxies with detected magnetic fields and those without detections. The possible exception is IRAS~F04332+0209, which is at the very low end of the sample in terms of $L_{\rm OH}$, $L_{\rm FIR}$, and $L_{\rm CO}$, such that some of the assumptions that we make about conditions in (U)LIRGs may not apply. In particular, we do not apply the inclination correction to Equation \ref{eq:dyn_mass} for IRAS~F04332+0209, as its CO luminosity is inconsistent with such a large dynamical mass. Nevertheless, using the method from S97 to estimate the gas surface density produced an answer consistent with the limits on gas surface density from the CARMA observations. The galaxy also shows evidence for extremely strong magnetic fields, as at least one pair of lines in its OH spectrum appear to be completely separated by Zeeman splitting, by an amount corresponding to a 46~mG field. This result requires confirmation with interferometric observations, however. We do include this galaxy in the remainder of the paper, but the overall results are unaffected if we exclude it.

Galaxies that also had detections of the CO(1-0) line in the literature were
then selected, and the procedure from S97 used to estimate gas surface densities and number densities. We expect the procedure is applicable to our sample as a whole, as the mean $L_{\rm FIR}$ and $L_{\rm CO}$ for our sample are both within 10\% of the values for Arp~220. 
Three additional OHMs with Zeeman detections or constraints were detected in the CO(1-0) line using CARMA (one of which was completely spatially unresolved). We use the results of the CARMA observations here, but defer more detailed presentation of the data to a later paper. These data, fluxes at 1.4~GHz, 60~$\mu$m, and 100~$\mu$m, and values derived from these measurements are shown in Table \ref{tab:data}.

{\em To estimate $\bism$ from $\boh$, we adopt a density contrast between OH masing regions and the mean ISM density of 10.} Based on modeling of OHM excitation, masing region with densities of $\nhtwo \sim 10^4 \cmc$ are successful at reproducing the general characteristics of the OHM population, and thermalization of OH precludes densities significantly above $10^5 \cmc$ \citep[M. Elitzur, private communication;][]{Elitzur1992book,Parra2005a}, while significantly lower gas densities are unlikely to be able to produce sufficient column densities for masing. For this reason, taking $\nhtwo \sim 10^4 \cmc$ is appropriate for masing clouds generally, though there is likely significant scatter. Though we made individual estimates of the ISM densities in each galaxy based on the S97 procedure, these values are also likely uncertain by a factor of a few in either direction, so we simply take the median, $\nhtwo \sim 10^3 \cmc$. This value is at the low end of density estimates made in Arp~220, but is consistent with the low density phase of molecular gas in its ISM. Given these values, a typical density contrast of 10 between OH masing regions and the ambient ISM appears reasonable. 

The final step is to use the density contrast to scale $\boh$ to $\bism$. Within the Milky Way, analysis of Zeeman splitting results finds that $B \sim B_0$ for densities below $n \sim 300 \cmc$, where $B_0$ is a constant, and $B \propto n^{0.65}$ at larger densities \citep{Crutcher2010}. These represent two extreme theoretical cases, the first corresponding to mass accumulating along field lines, and the second to spherical contraction unaffected by the magnetic field. \citet{Crutcher2010} interpreted the transition density as the point at which clouds become self gravitating. Even if a similar relationship holds in starburst galaxies, it is not obvious to which of these regimes masing clouds in starburst galaxies belong. The modeling of masing in III~Zw~35 by \citet{Parra2005a} concluded that individual masing clouds are not gravitationally bound. We take a compromise between these extreme possibilities, and adopt $B \propto n^{1/2}$, which roughly corresponds to $\bism \sim \boh / 3$ for the assumed density contrast of $n_{\rm OH}({\rm H}_2) \sim 10 n_{\rm ISM}({\rm H}_2)$.

\section{Results} \label{sec:results}

\begin{figure*}
    \centering
    \includegraphics[width=7in] {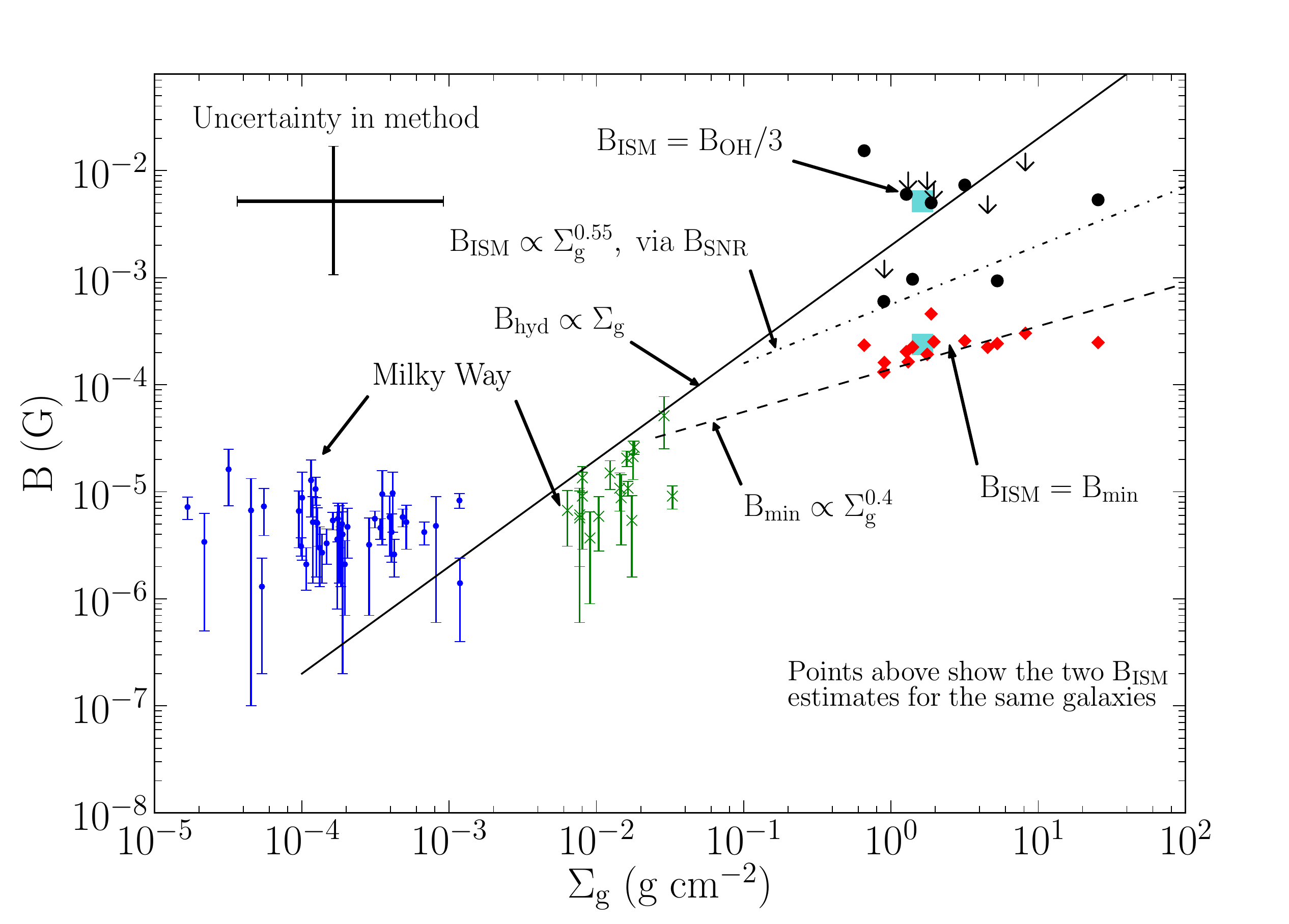}
    \caption{A variety of magnetic field measurements are shown as a function of the gas surface density. Magnetic fields within the Milky Way include Zeeman splitting measurements in non-gravitating HI clouds (blue dots with errorbars, from \citealt{Heiles2004}) and in OH toward dark cloud cores (green crosses with errorbars, from \citealt{Troland2008}). Two sets of magnetic field measurements are shown for starburst galaxies: magnetic field strengths inferred from Zeeman splitting in OH maser lines (black circles for detections, black downward pointing arrows for upper limits), and the minimum energy magnetic field for the same galaxies (red diamonds). For each magnetic field estimation method, the median $\Sigma_g$ and $B$ of galaxies with Zeeman detections is also shown (cyan squares). Finally, the expected relationship between $\Sigma_g$ and $B$ for three different cases is shown: a magnetic field in equipartition with gravity (solid line), the magnetic field if the minimum energy argument is correct (dashed line), and the magnetic fields inferred from observations of SNRs by \citet{Thompson2009} (dash-dot line).} \label{fig:mag_surfdens}
\end{figure*}

In Figure \ref{fig:mag_surfdens}, we directly compare the scaled Zeeman splitting measurements of magnetic field strength to that derived from synchrotron observations and the minimum energy argument. The plotted $\bmin$ values were derived following the assumptions used in T06, with a synchrotron volume filling factor of 1 and an energy ratio of 100 to 1 between cosmic ray ions and cosmic ray electrons. We further add two samples of Zeeman splitting derived magnetic field strengths from within the Milky Way. At gas surface densities of $\Sigma_g \sim 10^{-5}$--$10^{-3} \gcms$, magnetic fields measurements are Zeeman splitting of HI in sheets that are not likely to be self gravitating \citep{Heiles2004}. The measurements at $\Sigma_g \sim 10^{-2}$--$10^{-1} \gcms$ are from OH towards dark clouds \citep{Troland2008}. 

Three lines are also drawn in Figure \ref{fig:mag_surfdens}. The solid line shows $\beq$, the magnetic field strength if its energy density is equal to the pressure of a self gravitating gas disk in hydrostatic equilibrium, as a function of gas surface density. The dashed line shows a semi-empirical relationship for $\bmin$, based on magnetic field strengths derived from synchrotron fluxes and gas surface density, and was provided in Section 4.1 of T06. It takes the form $\bmin \propto \Sigma_g^{2/5}$ as result of a Kennicutt-Schmidt star formation law with a power law index of 7/5 \citep{Kennicutt1998}. Finally, the dash-dot line shows magnetic fields derived from a least squares fit of the radio luminosities of supernova remnants in starburst galaxies by \citet{Thompson2009}. The fit is to the magnetic field in the ISM as derived using observations of supernova remnants (SNRs), $\bsnr$. In brief, their argument is that the shock compressed ISM magnetic field is a lower limit to the field in SNRs, and thus observations of synchrotron emission from SNRs can be used to provide an upper limit to the ISM field. There is some uncertainty in this constraint because of uncertainty in when during the evolution of a SNR most of the relativistic electron acceleration occurs.

We add to this diagram our estimate of $\bism$ ($\boh / 3$) for galaxies with Zeeman detections and upper limits. As the uncertainty in converting the magnetic field strength in OH masing regions to that in the ISM as a whole dominates the statistical uncertainty associated with the measurements, we do not provide error bars for individual points. Instead, there is a single set of error bars in the upper left that roughly reflect the uncertainty in the methods used for estimating $\Sigma_g$ and $\bism$. We also added two cyan squares, which show the median $\bism$ and $\bmin$ of all galaxies with Zeeman detections at the location of the median $\Sigma_g$.  

We do not expect that the median is strongly affected by selection bias. \citet{McBride2013} detected Zeeman splitting in 14 galaxies, and placed upper limits on Zeeman splitting in an additional 26 galaxies. The median magnetic field detected was 12~mG. Only two galaxies (IRAS~F06487+2208 and IRAS~F12112+0305) had upper limits on the magnetic field associated with a masing line that were less than 12~mG, and two more had upper limits between 10--20~mG. Three galaxies, all in the sample presented here, had median field strengths $\sim$2--3 mG. Additionally, Zeeman splitting measures the magnetic field along the line of sight, $B_\parallel$, rather than the total magnetic field strength, except in cases where the splitting is greater than the linewidth. The marginal increase in the median as a result of more easily detecting large fields is thus largely offset by not applying any geometrical correction, and so the median should be representative.

The median magnetic field measured via Zeeman splitting ($\boh$) in OHMs is a factor of 60 larger than the median minimum energy magnetic field, with the ratio of the two varying from roughly 10 to 100. We have argued that $\bism \sim \boh / 3$, which then suggests that $\bism \sim 20 \bmin$ in starburst galaxies. There is significant uncertainty in this estimate, and it is based on a limited number of detections. Nevertheless, even if the magnetic field disparity between masing clouds and the ISM in starbursts is a factor of a few larger than we estimate, $\bism$ would be significantly larger than $\bmin$. Assuming the same scaling between magnetic field and density $B \propto n^{1/2}$, and the median ISM density $\nhtwo \sim 10^3 \cmc$, the density of masing regions implied by $\bism = \bmin$ would approach $\sim 10^7 \cmc$. This is above the density $10^5$--$10^6 \cmc$ at which OH thermalizes. Alternatively, if the masing density is $10^4 \cmc$, then $\bism \sim \bmin$ could occur if the volume averaged density in the region where synchrotron is produced is $\nhtwo \sim 1 \cmc$, given the ratio of $\boh$ to $\bmin$. This too seems unlikely for a galaxy like Arp~220, as its synchrotron emission occurs in a volume with radius $\sim$100~pc, where the diffuse molecular gas (not bound in clouds) has a density $\nhtwo \sim 10^3$--$10^4 \cmc$. Finally, if we vary the scaling between magnetic field and density we use to be $B \propto n^{2/3}$, the derived values of $\bism$ are smaller, but still consistent with $\bism \sim 10 \bmin$ for the adopted density contrast of 10 between OH masing clouds and the starburst ISM. Thus, the evidence from OHMs argues that it is quite unlikely that $\bism \sim \bmin$ in starburst galaxies.

\begin{figure}
    \centering
    \includegraphics[width=3.5in] {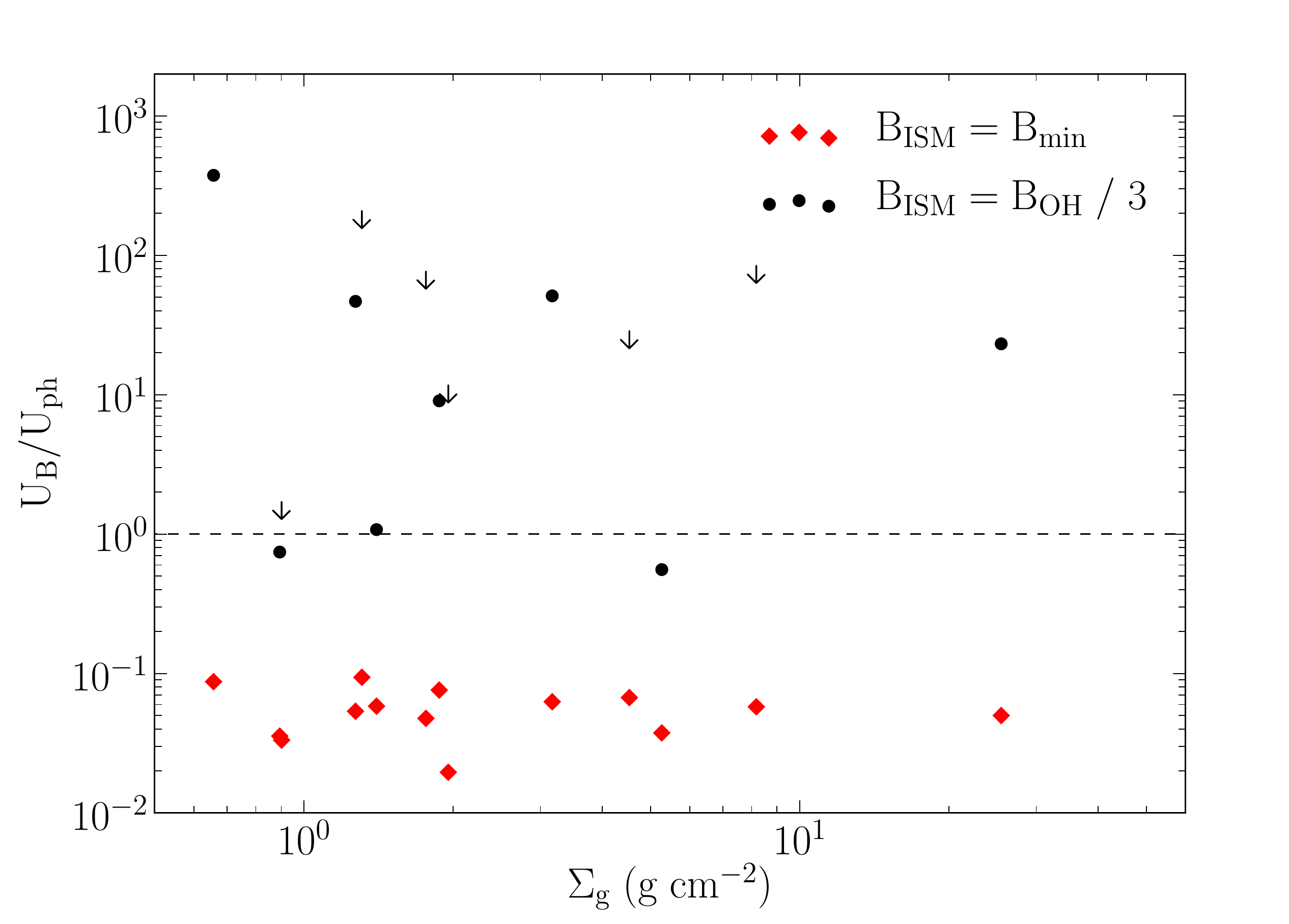}
    \caption{A dashed line marks equality between the magnetic and photon energy densities. For the FIR-radio correlation to hold in starburst galaxies, cosmic rays are expected lose their energy primarily via synchrotron emission, rather than inverse Compton. In this case, the magnetic energy density should be greater than the photon energy density, lying above the dashed line. Magnetic fields derived from OHMs (black circles, and limits denoted by black arrows) lie above the dashed line, while magnetic fields derived from synchrotron fluxes and the minimum energy argument (red diamonds) lie systematically below the dashed line.} \label{fig:energy_density}
\end{figure}

The median magnetic field derived from all Zeeman detections also suggests that $\bism \sim \beq$. Given the uncertainty in the data and the assumptions we used, we cannot robustly claim that $\bism \sim \beq$ in starburst galaxy ISMs, though our results are consistent with that possibility in at least a fraction of starburst galaxies. That $\bism \sim \beq$ appears initially to be at odds with the conclusion of \citet{Thompson2009}, who note that their results in high surface density galaxies imply that $\bism \ll \beq$. Figure \ref{fig:mag_surfdens} shows this result, as their magnetic field estimate is represented by the line $B \propto \Sigma_g^{0.55}$. However, this conclusion is dominated by the two nuclei of Arp~220, for each of which \citet{Thompson2009} estimate $\bism \sim 1$~mG. Their value of $\bism$ is in excellent agreement with our own estimate of $\bism$ from OH maser lines in that galaxy. The only other galaxy in their sample with $\Sigma_g > 1 \gcms$ is Arp~299. Our sample is almost entirely composed of galaxies with $\Sigma_g > 1 \gcms$, and has a median $\bism$ that is comparable to $\beq$. Thus the apparent tension of our result and that of \citet{Thompson2009} in Figure \ref{fig:mag_surfdens} may be entirely a consequence of Arp~220 having a weaker magnetic field than other (U)LIRGs, and the paucity of SNR observations in high gas surface density galaxies. If correct, then the clear prediction is that SNRs will be systematically brighter in other (U)LIRGs than in Arp 220. 

However, it is important to note that the utility of SNRs in constraining mean ISM magnetic fields has not been definitively established. For example, \citet{Batejat2011} used higher resolution observations of SNRs in Arp~220 to conclude that the SNR magnetic fields in starburst are unlikely to be caused by compression of the ISM field, as \citet{Thompson2009} concluded. \citet{Chomiuk2009} also argued that SNR magnetic field strengths are largely independent of ISM conditions, and that the conclusions of \citet{Thompson2009} were based on statistical sampling effects.

In evaluating evidence for $\bism > \bmin$ in starburst galaxies, \citet{Thompson2006} also compared the energy density in magnetic fields, $U_B$, to that in photons, $U_{\rm ph}$. They noted that the linearity in the FIR-radio correlation suggests that the synchrotron cooling timescale must be shorter than the inverse Compton cooling timescale, which requires $U_B > U_{\rm ph}$. $U_B$ values derived from $\bmin$ suggested that $U_B < U_{\rm ph}$ in starburst galaxies, again arguing for $\bism > \bmin$. Figure \ref{fig:energy_density} shows the ratio of $U_B$ to $U_{\rm ph}$ for $\bism$ derived from scaling $\boh$, as well as the same ratio for $\bism = \bmin$. $U_{\rm ph}$ is calculated using the FIR fluxes in Table \ref{tab:data} and $R_{\rm CO}$ determined from the S97 procedure described in Section \ref{sec:solomon_data}. For the entire sample of galaxies, $U_{\rm ph} > 10 U_B$ if $\bism = \bmin$. When taking $\bism = \boh / 3$, the ratio of magnetic to photon energy density is greater than unity for all but two sources, IRAS~F10038--3338 and Arp~220, and in both cases it is by less than a factor of two. This result suggests that the true magnetic energy density in starburst galaxies is comparable to or larger than the photon energy density. 

\section{Discussion} \label{sec:discussion}
T06 argued that in order for starburst galaxies to fall on the FIR-radio correlation, as they are observed to do, the magnetic fields should be sufficiently strong to provide synchrotron cooling times, $\tau_{\rm syn}$, that are less than the escape time for cosmic rays, $\tau_{\rm esc}$. The magnetic fields should also be strong enough so that the synchrotron cooling time will be shorter than the inverse Compton cooling time, $\tau_{\rm IC}$. For the magnetic fields strengths derived from Zeeman splitting, both criteria are met.

The synchrotron cooling time scales as $\tau_{\rm syn} \propto B^{-3/2}$. Using $\bism \sim 20 \bmin$ for the sample of galaxies considered here, the synchrotron cooling time in starbursts is smaller by roughly two orders of magnitude than the cooling time implied by $\bmin$. This is sufficient to produce $\tau_{\rm syn} < \tau_{\rm esc}$ for the values used in T06, which assumed the cosmic ray electron escape time was set by advection in a galactic wind. The result shown in Figure \ref{fig:energy_density} directly argues in favor of $\tau_{\rm syn} < \tau_{\rm IC}$, as $U_{B} \geq U_{\rm ph}$. Altogether, the magnetic field strengths inferred from Zeeman observations of OHMs provide evidence that cosmic ray electrons in starburst galaxies should radiate their energy primarily via synchrotron emission. This conclusion is consistent with the continued linearity of the FIR-radio correlation in starburst galaxies (T06) and the calorimeter theory of the FIR-radio correlation \citep{Voelk1989}.

The values of $\bism$ derived from $\boh$ are also consistent with $\bism \sim \beq$ in starburst galaxies, though this conclusion is by no means definitive. Equipartition magnetic field strengths would be expected if dynamo amplification has had sufficient time to strengthen the magnetic field. 
If dynamo processes have acted, then there may also be evidence for large scale structure in the magnetic field. Zeeman splitting in OHMs that also have VLBI maps already provide tentative evidence for large-scale magnetic field coherence. In III~Zw~35, for which we find $\bism \sim \beq / 3$, \citet{Robishaw2008} saw evidence for a magnetic field reversal on opposite sides of the $R \sim 20$\;pc torus in which masing occurs. They matched velocities of lines with Zeeman splitting to VLBI maps of the maser emission, and saw that the magnetic field in the north points away from us, and in the south it is pointed towards us. 

\citet{Robishaw2008} also looked for evidence of reversals in Arp~220, for which $\bism \sim \beq / 4.5$. In the western nucleus, there was no indication of field reversal. In the eastern nucleus, the data were ambiguous, as the velocity of some components allowed possible association with different spatial regions. One interpretation was consistent with reversal in the eastern nucleus. Resolving this ambiguity will require Zeeman observations using VLBI, rather than a single dish. 

Our conclusions about the mean ISM magnetic fields in (U)LIRGs depend upon the assumptions we outlined in Section \ref{sec:data}. The key assumptions are that the typical masing cloud in which Zeeman splitting is measured has a density of $\nhtwo \sim 10^4\cmc$, while the representative density for the majority of the volume of the starburst region in (U)LIRGs is $\nhtwo \sim 10^3\cmc$. These correspond to a density contrast of 10 between masing region and ISM, which we then use to scale $\boh$ to $\bism$ assuming that $B \propto n^{1/2}$. While we made all of these choices based on what appears to be the most reasonable interpretation of available data, there are multiple places where this estimate could be uncertain. 

For instance, OH masers associated with star formation in the Milky Way are thought to be produced in shocks around ultra-compact HII regions \citep{Fish2006}, and trace regions of higher density and amplified magnetic fields. Galactic OH masers and OHMs differ in many respects, however, and models of OHMs that explain many of their general features do not require masing produced in shocked material \citep{Parra2005a, Lockett2008}.  In addition, if the gas density of material in the synchrotron emitting region is significantly lower than we have assumed, that would also lead to our density contrast being an underestimate.  In a supersonically turbulent ISM, the gas density associated with (say) half the volume can be significantly smaller than the volume averaged gas density.  We have attempted to take this into account when choosing a density for the synchrotron emitting region that is a factor of $\sim 10$ smaller than the volume averaged gas density in the nuclear disks in Arp 220.  Future higher spatial resolution observations of many CO lines and higher density tracers in a large sample of starburst galaxies with the Atacama Large Millimeter Array will greatly improve our understanding of the molecular gas and ISM in (U)LIRGs \citep{Papadopoulos2012a}, providing an important check on the validity of our assumptions.

Arp~220 is the lone source with estimates of $\bism$ from both OH lines and from SNRs. The estimates agree very well, providing an important self consistency check for both methods. A more thorough consistency check for the two methods would be to make sensitive, high resolution observations capable of resolving individual SNRs in galaxies other than Arp~220 that also have detections of Zeeman splitting in OH maser lines. If the arguments outlined in this paper are correct, SNRs should be brighter in galaxies with larger values of $\bism$ as inferred from Zeeman splitting in OH maser lines. 

\section{Conclusions} \label{sec:conclusions}
The magnetic fields derived from Zeeman splitting in OH masing regions within (U)LIRGs argue strongly in favor of ISM magnetic field strengths that are systematically larger than those estimated by measurements of synchrotron flux and application of the minimum energy argument. While there are a number of assumptions and uncertainties underlying this conclusion, the conditions required in OH masing regions and the ISM to bring the two methods of estimating ISM magnetic field strength into agreement are implausible. A direct consequence of this conclusion is that the assumptions made in the minimum energy argument do not hold in the centers of starburst galaxies. 

The magnetic energy densities in starburst galaxies implied by the OHM derived magnetic field strengths are greater than the photon energy densities, meaning synchrotron cooling will dominate inverse Compton cooling. The magnetic fields are also strong enough that cosmic ray electrons should be expected to radiate their energy via synchrotron emission before escaping the galaxy. T06 argued that because starburst galaxies fall on the FIR-radio correlation, the synchrotron cooling time in starbursts must be shorter than the inverse Compton cooling time and the escape time for relativistic electrons. Our findings support T06's arguments for the ``calorimeter'' theory of the FIR-radio correlation \citep{Voelk1989}. \\

We thank the anonymous referee for providing useful suggestions that helped clarify the paper. We also thank Christopher McKee for reading the text and providing comments that improved the paper, and Moshe Elitzur, Leo Blitz, Tim Robishaw, Charles Hull, and Michael McCourt for useful conversations. 
This work was partially supported by a Simons Investigator award from the Simons Foundation to E. Q., the David and Lucile Packard Foundation, and the Thomas and Alison Schneider Chair in Physics at UC Berkeley. J. M. received support from a National Science Foundation Graduate Research Fellowship. We thank CARMA staff and observers for assistance in obtaining data. Support for CARMA construction was derived from the states of California, Illinois, and Maryland, the James S. McDonnell Foundation, the Gordon and Betty Moore Foundation, the Kenneth T. and Eileen L. Norris Foundation, the University of Chicago, the Associates of the California Institute of Technology, and the National Science Foundation. Ongoing CARMA development and operations are supported by the National Science Foundation under a cooperative agreement, and by the CARMA partner universities. This research used NASA's Astrophysics Data System Bibliographic Services, the SIMBAD database, operated at CDS, Strasbourg, France, and the NASA/IPAC Extragalactic Database (NED), which is operated by the Jet Propulsion Laboratory, California Institute of Technology, under contract with the National Aeronautics and Space Administration. Cosmological calculations used CosmoloPy (\href{http://roban.github.com/CosmoloPy/}{http://roban.github.com/CosmoloPy/}), a cosmology package for Python.

\bibliographystyle{apj}
\bibliography{ohm,sfg,extra,ism,radio}
\end{document}